\documentclass[aps,preprint]{revtex4}
\usepackage[dvips]{graphicx}
\usepackage{amssymb}
\usepackage{amsmath}
\usepackage{color}
\newcommand{\txtr}{\textcolor{red}}
\usepackage[normalem]{ulem}
\usepackage{gensymb}
\newcommand{\soutr}{\bgroup\markoverwith{\textcolor{red}{\rule[.5ex]{2pt}{1pt}}}\ULon}
\usepackage{natbib}
\usepackage{cancel}
\begin{document}

\draft
\title{Quantum graphs and microwave networks as narrow band filters for quantum and microwave devices}
\author {Afshin Akhshani, Ma{\l}gorzata Bia{\l}ous, and Leszek Sirko}
\address{Institute of Physics, Polish Academy of Sciences, Aleja Lotnik\'{o}w 32/46, 02-668 Warszawa, Poland}

\date{\today}

\bigskip

\begin{abstract}
	
We investigate properties of the transmission amplitude of quantum graphs and microwave networks composed of regular polygons such as triangles and squares. We show that for the graphs composed of regular polygons with the edges of the length $l$ the transmission amplitude displays a band of transmission suppression with some narrow peaks of full transmission. The peaks are distributed symmetrically with respect to the symmetry axis $kl=\pi$, where $k$ is the wave vector. For microwave networks the transmission peak amplitudes are reduced and their symmetry is broken due to the influence of internal absorption. We demonstrate that for the graphs composed of the same polygons but separated by the edges of length $l' < l$  the transmission spectrum is generally not symmetric according to the axis  $kl'=\pi$. We also show that graphs composed of regular polygons of different size with the edges being irrational numbers are not fully chaotic and their level spacing distribution and the spectral rigidity are well described by the Berry-Robnik distributions. Moreover, the transmission spectrum of such a graph displays peaks which are very close to $1$. Furthermore, the microwave networks are investigated in the time-domain using short Gaussian pulses. In this case the delay-time distributions, though very sensitive to the internal structure of the networks, show the sequences of transmitted peaks with the amplitudes much smaller than the input one. The analyzed properties of the graphs and networks suggest that they can be effectively used to manipulate quantum and wave transport.

\end{abstract}

\pacs{03.65.Nk,05.45.Mt}
\bigskip
\maketitle

% {\b {Introduction}}
\smallskip

\section{Introduction}

Quantum graphs of connected one-dimensional quantum wires were introduced by Pauling \cite{Pauling}.  Nowadays they play crucial role in the study of quantum chaos and complex quantum systems  \cite{Sanchez1998,Imry2002,Exner88,Kottos1999, Blumel2002,Hul2004,Berkolaiko2012,Pluhar2014,Dietz2017,Pinheiro2020,Lawrie2022}. A quantum graph is a metric graph $\Gamma=(V,E)$ composed of $v$ vertices, $v \in  V$, connected by one-dimensional $e$ edges, $e \in E$, which is equipped with the self-adjoint Laplace operator $L(\Gamma) = -\frac{d^2}{dx^2}$ acting in the Hilbert space of square integrable functions. It has discrete and non-negative spectrum \cite{Berkolaiko2012}.

The wave transport through a graph can be characterized by the off-diagonal elements $S_{12}(\nu)$ and $S_{21}(\nu)$ of the two-port scattering matrix $\hat S(\nu)$ of the graph \cite{Kumar2013,Lawniczak2020b}

\begin{equation}
\label{Eq.1}
\hat{S}(\nu)=\left[
\begin{array}{c c}
S_{11}(\nu) & S_{12}(\nu)\\
S_{21}(\nu) & S_{22}(\nu)
\end{array}
\right].
\end{equation}\

 Quantum graphs can be experimentally simulated by microwave networks  \cite{Hul2004,Lawniczak2008,Hul2012,Sirko2016,Dietz2017,Lawniczak2019b,Yunko2020}. Such simulation is possible because of the formal analogy of the one-dimensional Schr\"odinger equation describing quantum graphs and the telegrapher's equation describing microwave networks \cite{Hul2004,Sirko2016}.
 Microwave networks allow for the experimental realization of systems described by the main three symmetry classes in random-matrix theory (RMT): Gaussian orthogonal ensemble (GOE) \cite{Hul2004,Hul2005b,Lawniczak2008,Hul2012,Dietz2017,LipovskyJPhysA2016,Lawniczak2019,Lawniczak2020,Lawniczak2021c},  Gaussian unitary ensemble (GUE) \cite{Hul2004,Lawniczak2009,Lawniczak2010,Lawniczak2012,Allgaier2014,Bialous2016,Lawniczak2019b}  - systems with broken time reversal symmetry, and Gaussian symplectic ensemble (GSE) \cite{Stockmann2016,Dietz2021,Lawniczak2023}. Thus any theoretical prediction obtained for quantum graphs can be verified experimentally using microwave networks. Quantum graphs and microwave networks with preserved time reversal symmetry are reciprocal therefore their scattering matrices $\hat{S}(\nu)$ are symmetric which leads to the relationship: $S_{12}(\nu) = S_{21}(\nu)$. In this analysis we will consider the most frequently analyzed graphs and microwave networks with the Neumann, called also standard, vertex boundary conditions \cite{Hul2012} which impose that the functions are continuous at vertices and the sums of their oriented derivatives at vertices are equal zero.

For the completeness of the presentation one should mention that in the simulation of other complex quantum systems microwave plane billiards \cite{Dietz2010,Yeh2013, Zheng2006,Stockmann1990,Sridhar1994,Sirko1997,Hlushchuk2000,Hlushchuk2001,
	Hlushchuk2001b,Dhar2003,Savytskyy2004,HemmadyPRL2005,Hul2005,Dietz2015,Dietz2019} and atoms excited in strong microwave fields \cite{Jensen1991,Bellerman1992,Buchleitner1993,SirkoPRL1993,Bayfield1995,Sirko1995,Sirko1996,Bayfield1999,Sirko2001,Galagher2016} are used as model systems.

The transmission through a network accompanied by backscattering (reflection) of a wave is the main feature of particles and wave dynamics. The absence of backscattering leads to the  reflectionless absorption \cite{Anlage2020} or the reflectionless transmission \cite{Yusupov2019}. The latter case is particularly important from an application point of view, e.g. in the condensed matter while transport of spin, charge or other carriers are considered in nanomaterials.

The Green's function approach is often used in analyzing of scattering properties of quantum graphs \cite{Schmidt2003,Andrade2016,Andrade2018,Drinko2019,Drinko2020}. Drinko {\it et al.} \cite{Drinko2020} applied this mathematical formalism to investigate the scattering properties of dissipationless quantum graphs composed of regular polygons  such as triangles  $C_{3}$ and squares $C_{4}$ with the edges of the same optical length $l$ (see Fig.~1(a)). In Ref. \cite{Drinko2020} the global transmission amplitudes $t_{C_3}(k)$ and $t_{C_4}(k)$ were also calculated for the triangles  $C_{3}$ and squares $C_{4}$ possessing the Neumann vertex boundary conditions (Eq. (6) in \cite{Drinko2020}). However, they were printed with some errors. The correct formulas are the following

\begin{equation} \label{Eq.2}
\begin{split}
&t_{C_3}(k) = \frac{4z(z^3-1)(z+1)}{9-z^2-8z^3-z^4+z^6},\\
&t_{C_4}(k) = \frac{4z(z^4-1)(z^2+1)}{9-z^2-8z^4-z^6+z^8},
\end{split}
\end{equation}

where $z=e^{ikl}$.

One should also remark that in Ref. \cite{Estarellas2015} the quantum states of regular polygonal structures possessing from 3 to 6 vertices made of 1D quantum wires were calculated treating each polygon vertex as a scatterer.

In this article we go beyond a simple modeling of dissipationless graphs presented in Ref. \cite{Drinko2020} and analyze  numerically and experimentally open quantum graphs and microwave networks with internal dissipation and complicated geometry.

\section{Numerical calculations}

In the analysis of transport properties of graphs and microwave networks we will consider graphs and networks with the Neumann boundary conditions for which the vertex scattering matrices $\sigma_{e, e'}^{(v)}$  \cite{Farooq2022} are described by

\begin{equation} \label{Eq.3}
\sigma_{e, e'}^{(v)} = \frac{2}{d_v} - \delta_{e, e'},
\end{equation}

where $d_v$ is degree of the vertex $v$ and~$\delta_{e,e'}$ is the Kronecker~delta.

The numerical calculations of the transmission amplitude of graphs were performed using the method of pseudo-orbits \cite{LipovskyJPhysA2016,BHJ,Li7}.
In the calculations we took into account dissipation inside microwave networks by replacing the wave vector $k$, characterizing dissipationless systems, with the one $k+i\beta \sqrt{|k|}$ accounting  for their absorption,
where $\beta = 0.009 $ m$^{-1/2}$ \cite{Hul2004,Goubau1961}.

\txtr {The presence of dissipation in quantum graphs gives rise to a multitude of distinctive behaviors. It induces exponential decay of quantum states, leading to a rapid decrease in the probability amplitude over time.
Because of internal dissipation of the networks their resonances show up as poles \cite{Kottos2003,Lawniczak2019} occurring at complex wave numbers $k_m=\frac{2\pi}{c}(\nu_m -i\Delta\nu_m)$, where c is the vacuum velocity of light and $\nu_m$ and $2\Delta\nu_m$ are associated with the positions and the widths of resonances, respectively. Recently, the dissipative spectral form factor has been introduced to diagnose dissipative quantum chaos and reveal correlations between real and imaginary parts of the complex
eigenvalues \cite{Li2021}.
Dissipation disrupts interference patterns, a fundamental aspect of quantum mechanics, and dampens resonance behavior. This deteriorating influence of dissipation is especially important in application of quantum filters therefore we decided to study it experimentally and numerically.}

\section{Experiment}

In order to simulate quantum graphs we applied  microwave networks which are composed of microwave coaxial cables and $T$-junctions that correspond to the edges and vertices of degree three of quantum graphs, respectively. The coaxial cable consists of an outer conductor with an inner radius $r_{2} = 0.15$ cm and an inner conductor of a radius $r_{1} = 0.05$ cm. These two conductors are separated by  Teflon with an experimentally determined dielectric constant $\varepsilon=2.06$.
Below the cut-off frequency $\upsilon_{cut}=\frac{c}{\pi(r_{1}+r_{2})\sqrt{\varepsilon}} \simeq 33$ GHz \cite{Jones1964,Savytskyy2001} only the fundamental TEM mode propagates in a cable. The internal absorption of microwave networks was taken into account on the basis of absorption of their edges made of the microwave cables and their absorption coefficient $\beta = 0.009 $ m$^{-1/2}$.

The experimental investigations of the transport properties were realized by using microwave networks containing regular polygons: triangles $C_{3}$ and squares $C_{4}$. Each polygon is constructed of edges of the same optical length $l= \sqrt{\varepsilon}l_{ph} = 0.25$ m, where $l_{ph}$ is the physical length of the edge. In our investigations we focused on  microwave networks simulating quantum graphs containing 3 polygons: triangle-square-triangle $C_{3}C_{4}C_{3}$ and square-triangle-square $C_{4}C_{3}C_{4}$ (Fig.~1(a) and Fig.~1(b)). The polygons were connected by the edges of  length $l'$.  In the case of triangles $C_{3}$ and squares $C_{4}$ the edges of polygons connected to the vertices of degree $d_v=2$, where $d_v$ is the number of edges incident to the vertex $v$, were replaced by the rings of coaxial cables with the corresponding optical lengths $2l$ and $3l$ (see Fig.~1(c)).

The two-port scattering matrices $\hat S(\nu)$ of the microwave networks were measured  using Agilent E8364B vector network analyzer (VNA) (see Fig.~1(d)).  Two flexible microwave cables HP 85133-616 and HP 85133-617 were applied to connect the networks to the VNA. They are equivalent to attaching  semi-infinite external leads $\mathcal{L}_1^{\infty}$ and $\mathcal{L}_2^{\infty}$ to the quantum graph \cite{LawniczakAPPA2019}. In such a way the transmission amplitude $|S_{12}(\nu)|$ of the network was measured as a function of microwave frequency $\nu$.  In the real physical systems absorption of microwaves in coaxial cables is unavoidable. It results in resonance broadening and overlapping, therefore, the transmission spectra were measured for low frequency range $0.01-1.2$ GHz.

In Fig.~2(a) we present the transmission amplitude $T_{C_{3}C_{4}C_{3}}(k)\equiv |S_{12}(ck/2\pi)|$ as a function of $kl$ for the graph (filter)  $C_{3}C_{4}C_{3}$ (triangle-square-triangle) with the edges $l'=l$ (see Fig.~1(a)).  The numerical results (black solid line) obtained for the absorption coefficient $\beta = 0 $ m$^{-1/2}$ are in full agreement with the theoretical predictions utilizing the Green's function approach \cite{Drinko2020}.
The constructive quantum interference leads to the appearance of narrow peaks with the Full Width at Half Maximum $FWHM \simeq 3$ MHz (see  Fig.~2(a)). They achieve maximum of the transmission amplitude $T_{C_{3}C_{4}C_{3}}(kL)=1$ in the region of transport suppression.
The transmission amplitude obeys the relation $T_{C_{3}C_{4}C_{3}}(\pi+kl)=T_{C_{3}C_{4}C_{3}}(\pi-kl)$ for $kl\in[0,\pi]$.  In Fig.~2(a) we show also the experimental results (red solid line) obtained for the microwave network  simulating the quantum graph $C_{3}C_{4}C_{3}$. Due to internal absorption of the network the experimental transmission coefficient is compared to the numerical one  (green dotted line) obtained for the absorption coefficient $\beta=0.009$ m$^{-1/2}$. The agreement between these results are very good. Fig.~2(a) shows also that the experimental transmission peaks compared to the dissipationless situation are reduced by absorption leading to energy transport losses and the symmetry breaking in the transmission structures. It is important to point out that the amplitudes of the transmission peaks can be significantly increased by  reducing the length $l$ of the network edges.

In Fig.~2(b) we show the numerical results obtained for the unconsidered earlier dissipationless graph $C_{4}C_{3}C_{4}$ with $l'=l$ (black solid line). Also in this case  the transmission coefficient is symmetric with respect to the symmetry axis $kl=\pi$. In Fig.~2(b) we  compare the experimental results obtained for the microwave network $C_{4}C_{3}C_{4}$ (red solid line) and the numerical results obtained for the graph $C_{4}C_{3}C_{4}$ with the absorption coefficient $\beta=0.009$  m$^{-1/2}$  (green dotted line). Fig.~2(b) shows that both experimental and numerical results are in good agreement.

A completely different situation will appear if we change the construction of the filters $C_{3}C_{4}C_{3}$ and $C_{4}C_{3}C_{4}$ and allow the polygon elements of the graphs to be separated by the edges $l'<l$. With $l'$ rationally related to $l$, e.g. $l'=l/3$, we also expect to have multiple bands of transmission suppression with the narrow transmission peaks but this time distributed symmetrically according to the symmetry axis $kl'=\pi$. In Figs.~3(a) and 3(b) we show the numerical results obtained for the dissipationless graphs $C_{3}C_{4}C_{3}$ and $C_{4}C_{3}C_{4}$ with $l'=l/3$ (black solid lines). To show the symmetry in the transmission coefficient the range of $kl$ was extended to $6\pi$. Indeed, the symmetry axes of the transmission structures of the graphs at $kl=3\pi$ is clearly seen. For $l'=l/3$ the positions of the transmission peaks are differently distributed than in the case of $l'=l$, showing that the properties of the filters can be easily modified by replacing the edges $l'=l$ with the edges $l'$ rationally related to $l$. In Figs. 3(a) and 3(b) the transmission amplitude of the dissipationless graphs is compared with the one obtained for the absorption coefficient $\beta=0.009$ m$^{-1/2}$ (blue dotted line). Also in this case the presence of absorption causes that the transmission peaks are reduced in heights leading to energy transport losses and the symmetry breaking in the transmission structures.

For the graphs $C_{3}C_{4}C_{3}$ and $C_{4}C_{3}C_{4}$ with the edges $l'$ not rationally related to $l$ we do not have a symmetry axis in the transmission amplitude of the graphs. As an example of such a situation in Figs.~4(a) and 4(b) we show the transmission amplitude (black solid line) for the dissipationless filters $C_{3}C_{4}C_{3}$ and $C_{4}C_{3}C_{4}$ with $l'=l/\pi$.  Indeed in the range $0\leq kl \leq 6\pi$ the transmission amplitudes show the unique structures without any visible repetition. However, even in such a case the structures of the suppression bands with the narrow transmission peaks are also present and the tuning of $l'$ can be effectively used to modify the filters properties. The results for the  dissipationless graphs are compared to the ones obtained for the graphs with the absorption coefficient $\beta=0.009$  m$^{-1/2}$  (red dotted line), showing that the presence of absorption causes the transmission peaks reduction.

To consider even more general situation in this article we analyze the transmission spectrum of the graph $C_{3}C_{4}C'_{3}$ (see Fig.~5) composed of regular polygons of different size with edges that are not rationally related.  The edge lengths of the polygons are the following: $l/e$, $l/\sqrt{3}$, and $l/\sqrt{5}$ for the polygons $C_{3}$, $C_{4}$, and $C'_{3}$, respectively, where $e=2.71828182...$ is the Euler's irrational number. The polygons are separated by the edges of the length $l/\pi$. In Fig.~5(b) we show the transmission amplitude of the dissipationless graph (grey solid line) which is compared with the one obtained for the absorption coefficient $\beta=0.009$ m$^{-1/2}$ (green dotted line). Even in the case of the dissipationless graph not all resonances are well separated suggesting that we deal with a system characterized by the intermediate between Poisson and GOE  level spacing distribution.

To corroborate this observation we performed the analysis of the spectral statistics of the transmission spectra. For that reason the set of ordered eigenvalues of the graph was converted to a set of normalized spacing.  The procedure is carried out by replacing the resonance frequencies $\nu_{m}$ by the smooth part of the integrated level density, that is given by the Weyl's formula $N_{av}(\nu_m)$ \cite{Lawniczak2019}

\begin{equation}
	\label{Eq.4} \epsilon_{m} = N_{av}(\nu_m) \simeq \frac{2\mathcal{L}}{c}\nu_{m},
\end{equation}

where $\mathcal{L}$ is the sum of the lengths of the edges of the graph. This gives dimensionless eigenvalues $\epsilon_{m}$ with mean value unity, $\langle s\rangle=1$ of the spacing $s_{m}=\epsilon_{m+1}-\epsilon_{m}$ between adjacent levels. Properties of the spectral statistics will be analyzed applying the nearest-neighbor spacing distribution (NNSD) and the spectral rigidity. The NNSD is the most commonly used measure of spectral regularity of quantum systems which gives information on short-range correlations. The analytical results for the NNSD for classically regular systems $P^{Poisson}(s)$ and the chaotic ones described by the Gaussian Orthogonal Ensemble in RMT  $P^{GOE}(s)$ are given by the following formulas \cite{Mehta1990}

\begin{equation}
	\label{Eq.5} P^{Poisson}(s)=\mbox{e}^{-s},
\end{equation}
\begin{equation}
	\label{Eq.6} P^{GOE}(s) =\frac{\pi}{2}s\mbox{e}^{-(\pi/4)s^2}.
\end{equation}

 Spectral properties of complex systems that are usually not fully regular or fully chaotic that can be described by statistically independent superposition of the Poisson and GOE statistics are well approximated by the Berry-Robnik distribution $P^{BR}(s,\rho_1)$ \cite{Berry1984,Prosen1994}

\begin{equation}
	\label{Eq.7} P^{BR}(s,\rho_{1})=\rho_{1}^{2} \mbox{e}^{-\rho{_1}s}\mbox{erfc}(\frac{1}{2}\sqrt{\pi}\rho_{2} s) +(2\rho_{1}\rho_{2} + \frac{1}{2}\pi\rho_{2}^{3}s)\exp(-\rho_{1}s-\frac{1}{4}\pi\rho_{2}^{2}s^{2}),
\end{equation}	

where $\mbox{erfc}(\cdot)$ is the complementary error function and $\rho_2=1-\rho_1$.

It interpolates between the Poisson $P^{Poisson}(s)$ ($\rho_1=1$, $\rho_2=0$) and the Wigner distribution $P^{GOE}(s)$ ($\rho_1=0$, $\rho_2=1$), i.e. it characterizes the transition from fully uncorrelated to correlated energy-levels.
By fitting the Berry-Robnik distribution to NNSD  the parameter $\rho_1$ can be evaluated.

In Fig.~5(c) we show the NNSD (histogram) obtained using 1811 eigenfrequencies of the graph. The numerical NNSD is compared to the Poisson (green dashed line) and GOE (blue dash-dotted line) theoretical distributions.  The fit of the formula (\ref{Eq.7}) (solid red line) to the numerical data yields the parameter $\rho_1=0.496 \pm 0.019$  which shows that the numerical NNSD displays the intermediate quantum behavior characterized by the mixture of uncorrelated and correlated energy-levels. In Fig~5(c) the error bars represent the standard deviation of uncertainty.

The spectral rigidity $\Delta_3(L)$ measures the least square deviation of the spectral staircase function $N(\epsilon)$ from the best line fitting it

\begin{equation}
	\label{Eq.8} \Delta_3(L)=\frac{1}{L}\langle \mbox {min}_{A,B}\int_{-L/2}^{L/2}[N(\epsilon)-A\epsilon -B]^2 d\epsilon\rangle,
\end{equation}	

where $\langle \cdot \rangle$ denotes a local average.
The statistical independence of regular and chaotic components of the mean level density $\rho_1$ and $\rho_2$ leads to the theoretical formula for the Berry-Robnik spectral rigidity \cite{Prosen1994}

\begin{equation}
	\label{Eq.9} \Delta^{BR}_3(L,\rho_{1})= \Delta^{Poisson}_{3}(\rho{_1}L)+ \Delta^{GOE}_{3}(\rho{_2}L),
\end{equation}	

where $\Delta^{Poisson}_{3}(L)$ and $\Delta^{GOE}_{3}(L)$ are the spectral rigidity distributions for the Poisson and GOE systems, respectively.

In Fig.~5(d) the numerical spectral rigidity $\Delta_3(L)$ (grey open circles) evaluated for the graph $C_{3}C_{4}C'_{3}$ is compared to the Berry-Robnik spectral rigidity $\Delta^{BR}_3(L,\rho_{1})$ (solid red line), calculated for  $\rho_1=0.496 \pm 0.019$. The error bars in Fig~5(d)  represent the standard deviation of uncertainty. The agreement between the numerical spectral rigidity $\Delta_3(L)$ and the Berry-Robnik spectral rigidity $\Delta^{BR}_3(L,\rho_{1})$ is very good. In Fig~5(d) the numerical spectral rigidity $\Delta_3(L)$ is also compared to the spectral rigidity distributions $\Delta^{Poisson}_{3}(L)$ (green dashed line) and $\Delta^{GOE}_{3}(L)$ (blue dash-dotted line) predicted for classically regular and chaotic quantum systems, respectively.

Summarizing, the nearest-neighbor spacing distribution $P(s)$ and the spectral rigidity $\Delta_3(L)$ of the graph  $C_{3}C_{4}C'_{3}$  show the transition from Poisson to GOE quantum chaotic behavior which is well described by the Berry-Robnik distributions. It is important to point out that despite that the graph is partially chaotic its transmission spectrum displays peaks which are very close to $1$, e.g. at $kl \simeq 7.2\pi$ and $11.3\pi $,  allowing to use such a graph as a spectral filter.

Additionally to the continue wave measurements we also investigated  the scattering properties of the networks presented in Fig.~1(c) and Fig.~6(a) using  delay-time distributions of short Gaussian pulses (see Fig.~6).
The signal of $FWHM= 125$ ps and amplitude A=0.41 V was synthesized by the waveform generator AWG 7082C. The output signal after network's penetration was detected and stored by the digital oscilloscope MSO-X-91304A. In Fig.~6(b) we show the experimental delay-time distributions obtained for the microwave networks $C_{3}C_{4}C_{3}$ and $C_{4}C_{3}C_{4}$ with $l'=l$ denoted by blue dotted and green solid lines, respectively.  The positive and negative peaks transmitted through the networks are formed due to the interference  of signals traveling along many very complicated available paths. For example, the first 3 peaks recorded for the network $C_{3}C_{4}C_{3}$, marked as $5l$, $6l$, and $7l$, are attributed to the paths: $acdghj$, $abcdghj+acdghij$, and $acdefghj+abcdghij+acacdghj+acdcdghj+acdgdghj+acdghghj+acdghjhj$, respectively. In the case of the network $C_{4}C_{3}C_{4}$ the first 3 peaks $5l$, $6l$, and $7l$ are attributed to the paths: $adeghk$, $adefghk$, and $abcdeghk+adeghijk+adadeghk+adedeghk+adegeghk+adeghghk+adeghkhk$, respectively. The application of the Neumann boundary conditions (Eq.~\ref{Eq.3}) to vertices being on the paths assigned to the peaks $5l$, $6l$ and $7l$ shows that for both networks the peaks $5l$ and $7l$ should have the same heights while the peak $6l$ for the network  $C_{3}C_{4}C_{3}$ should be 2 times higher than the one for the network  $C_{4}C_{3}C_{4}$. Fig.~6(b) shows that the theoretical predictions are in agreement with the experimental results.  Our experimental results also show that the sequences of the transmitted peaks possess much smaller amplitudes than the input one. For example the ratio of the amplitude of the peak $7l$ and the input one for the dissipationless network should be $U^{7l}/U^{in}=0.224$, what is very close to the experimental one $0.21 \pm 0.02$.

\section{Summary}

We investigated numerically and experimentally transport properties of quantum graphs and microwave networks $C_{3}C_{4}C_{3}$ and $C_{4}C_{3}C_{4}$ composed of simple regular polygons: triangles $C_{3}$ and squares $C_{4}$. The transmission spectra exhibit suppression bands, i.e. regions of full transmission suppression with narrow peaks of $FWHM \simeq 3$ MHz and maximum transmission which are the consequence of constructive wave interference. The structures of bands and transmission maxima can be controlled by changing the geometry of the graphs. In particular if the polygon elements of the graphs are  separated by the edges $l'$ irrationally related to $l$ the transmission structures of the graphs do not posses any symmetry axis. The numerical results for the absorption-free graph $C_{3}C_{4}C_{3}$ are in good agreement with theoretical predictions \cite{Drinko2019}. We also investigated microwave networks simulating quantum graphs with absorption. We showed that the experimental results are in agreement with the numerical ones obtained for the graphs with the absorption coefficient $\beta=0.009$ m$^{-1/2}$.
The nearest-neighbor spacing distribution $P(s)$ and the spectral rigidity $\Delta_3(L)$ of the graph $C_{3}C_{4}C'_{3}$ composed of regular polygons of different size with the edges that are not rationally related show intermediate behavior between the Poisson and GOE one,  well described by the Berry-Robnik distributions. We showed that the transmission spectrum of such a graph displays transmission peaks which are very close to $1$, e.g. at $kl \simeq 7.2\pi$ and $11.3\pi $, indicating that even the graphs composed of irrational edges can be used as spectral filters.
The time resolved domain was experimentally investigated measuring  delay-time distributions of short Gaussian pulses of $FWHM= 125$ ps propagating through the  microwave network.
Our results shed new light on the design and construction of new spectral devices, in particular ultra-narrow bandpass filters, with controllable multiple transmission maxima and zeros, which can be used to manipulate the quantum and wave transport.

\section{Acknowledgments}

This work was supported by the National Science Centre, Poland, Grant No. 2018/30/Q/ST2/00324.

\pagebreak

%\centerline {\bf Figure Captions}

\smallskip

\begin{figure}[tb]
\begin{center}
\rotatebox{0}{\includegraphics[width=1.2\textwidth,
height=1.2\textheight, keepaspectratio]{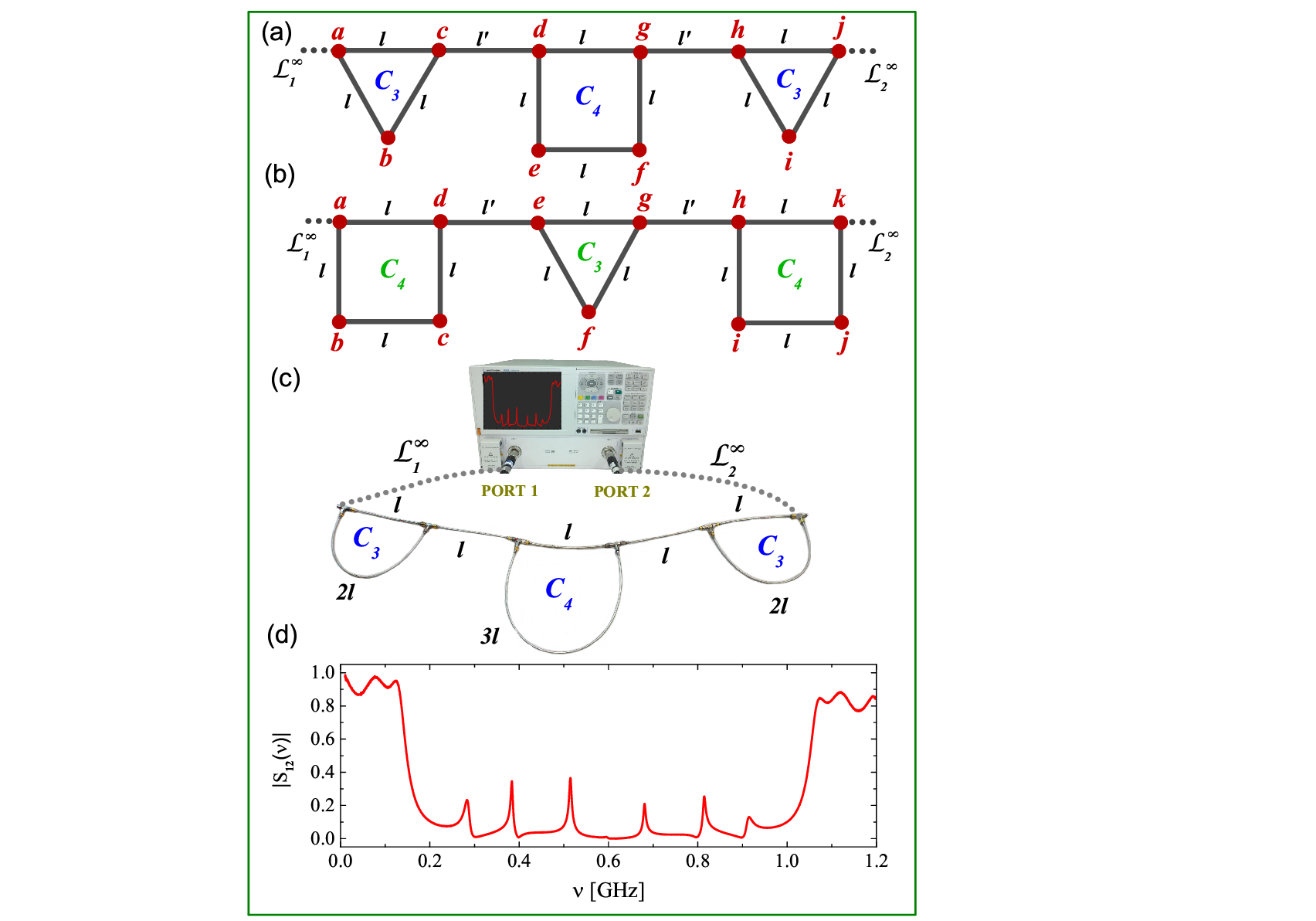}}
 \caption{The schemes of the triangle-square-triangle $C_{3}C_{4}C_{3}$ (panel (a)) and the square-triangle-square  $C_{4}C_{3}C_{4}$ (panel (b)) quantum graphs with vertices of degree 3 and 2.  (c) The photo of the microwave network $C_{3}C_{4}C_{3}$. The network was connected to the ports of the vector network analyzer Agilent E8364B in order to measure the transmission amplitude $|S_{12}(\nu)|$ of the network (panel (d)).
}\label{Fig1}
\end{center}
\end{figure}

\begin{figure}[tb]
\begin{center}
\rotatebox{0}{\includegraphics[width=1.2\textwidth,
height=1.2\textheight, keepaspectratio]{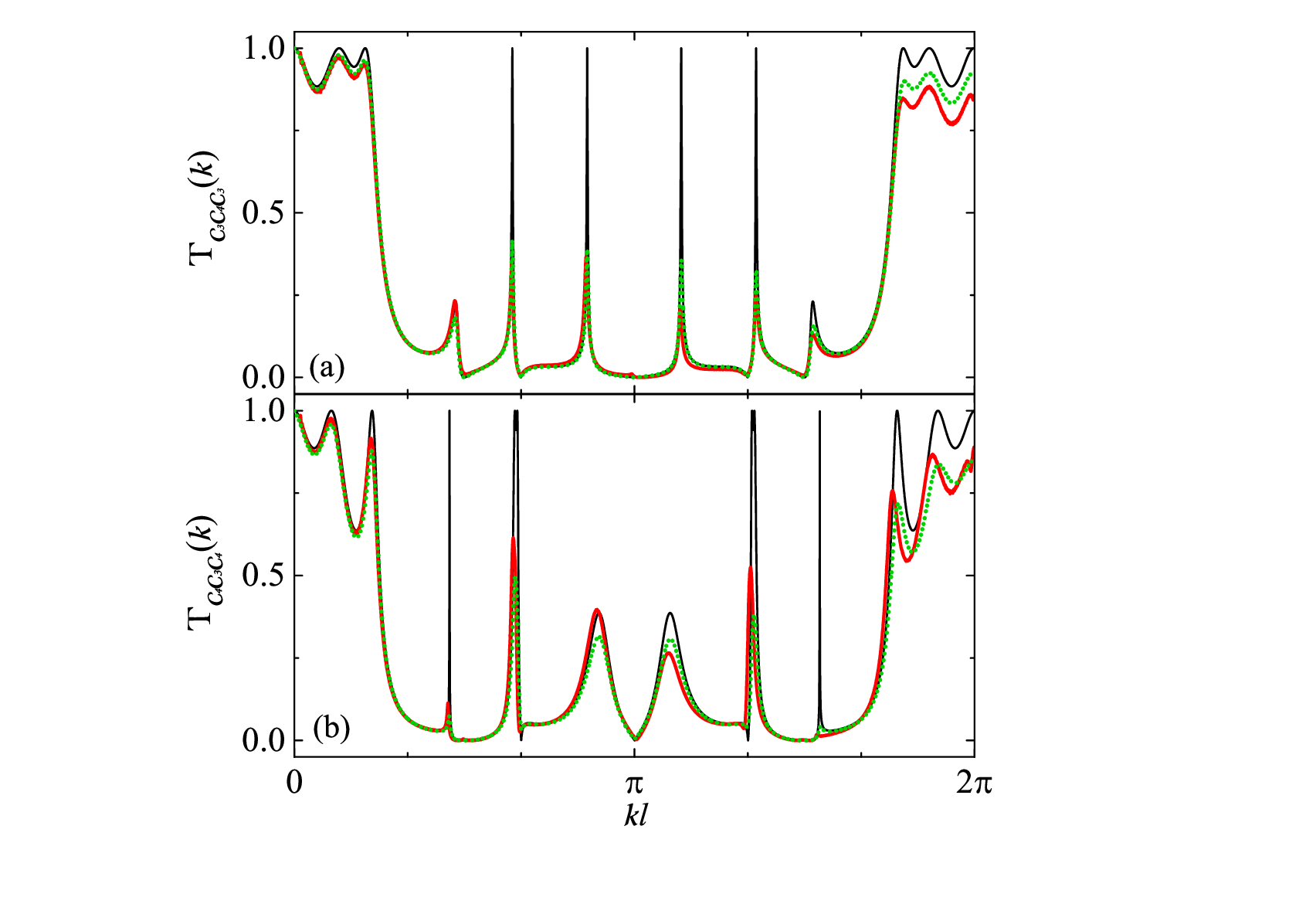}}
\caption{The transmission amplitude of the graphs and microwave networks composed of regular  triangles and squares with the edges of the same optical length $l$  separated by the edges of the length $l'=l$ (see Fig.~1): (a) the triangle-square-triangle graph $C_{3}C_{4}C_{3}$ and (b) the square-triangle-square graph $C_{4}C_{3}C_{4}$. The numerical results with the absorption coefficient $\beta=0$ and $\beta=0.009$ m$^{-1/2}$ are demonstrated respectively by black solid and green dotted lines. The experimental spectra of the microwave networks are shown by red solid lines.}\label{Fig2}
\end{center}
\end{figure}

\begin{figure}[tb]
\begin{center}
\rotatebox{0}{\includegraphics[width=1.2\textwidth,
height=1.2\textheight, keepaspectratio]{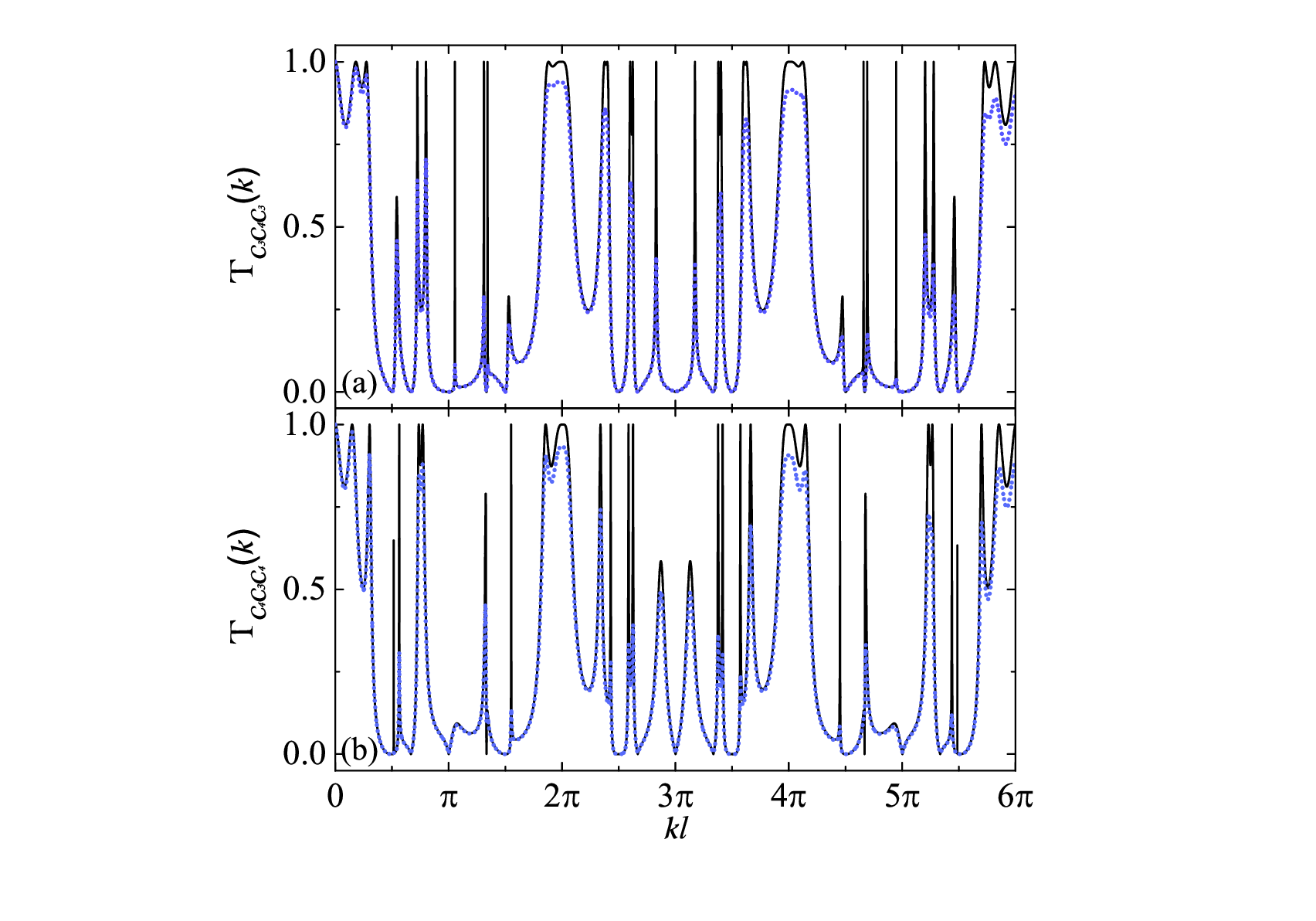}}
\caption{The transmission amplitude of the graphs composed of regular triangles and squares with the edges of the same optical length $l$  separated by the edges of the length $l'=l/3$ (see Fig.~1): (a) the triangle-square-triangle graph $C_{3}C_{4}C_{3}$ and (b) the square-triangle-square graph $C_{4}C_{3}C_{4}$. The numerical results with the absorption coefficient $\beta=0$ and $\beta=0.009$ m$^{-1/2}$ are shown by black solid and blue dotted lines, respectively.
}\label{Fig3}
\end{center}
\end{figure}

\begin{figure}[tb]
\begin{center}
\rotatebox{0}{\includegraphics[width=1.2\textwidth,
height=1.2\textheight, keepaspectratio]{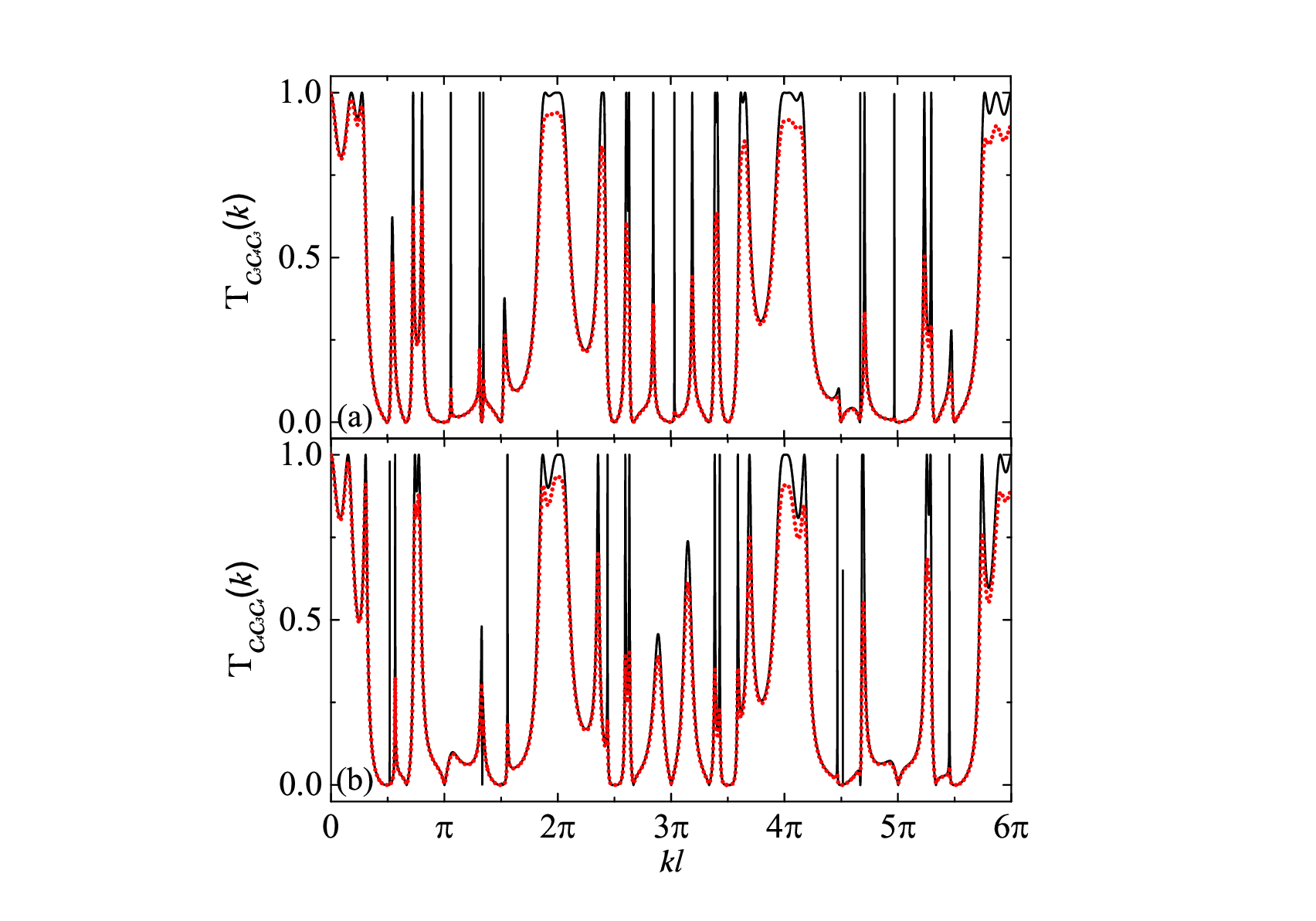}}
\caption{The transmission amplitude of the graphs composed of regular triangles and squares with the edges of the same optical length $l$  separated by the edges of the length $l'=l/\pi$ (see Fig.~1): (a) the triangle-square-triangle graph $C_{3}C_{4}C_{3}$ and (b) the square-triangle-square graph $C_{4}C_{3}C_{4}$. The numerical results with the absorption coefficient $\beta=0$ and $\beta=0.009$ m$^{-1/2}$ are shown by black solid and red-yellow dotted lines, respectively.
}\label{Fig4}	
\end{center}
\end{figure}

\begin{figure}[tb]
\begin{center}
\rotatebox{0}{\includegraphics[width=1.2\textwidth,
height=1.4\textheight, keepaspectratio]{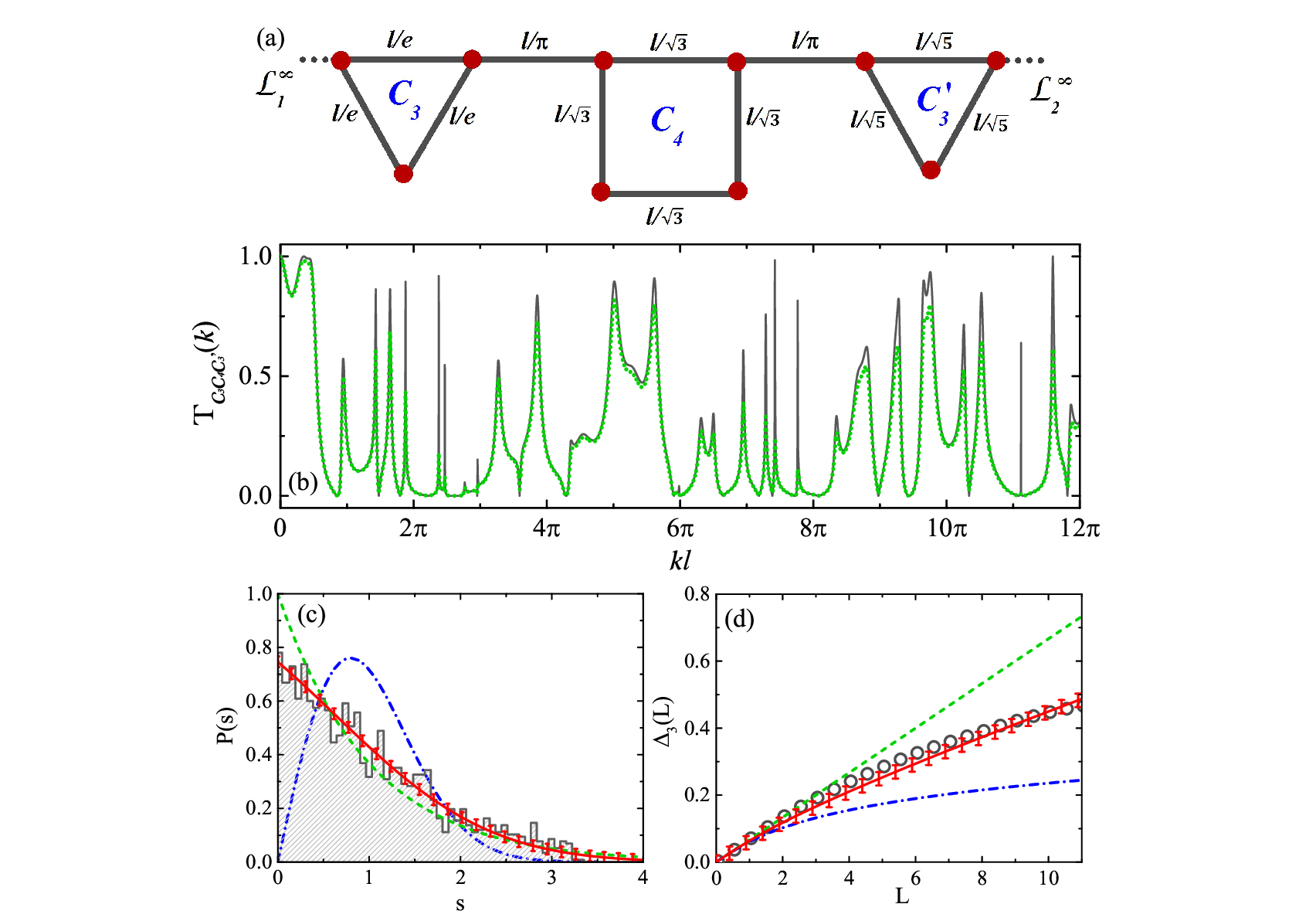}}
\caption{(a) The scheme of the graph $C_{3}C_{4}C'_{3}$ composed of regular polygons of different size with edges that are not rationally related. (b) The transmission amplitude of the dissipationless graph (grey solid line) is compared with the one obtained for the absorption coefficient $\beta=0.009$ m$^{-1/2}$ (green dotted line). (c) The nearest-neighbor spacing distribution (NNSD) (histogram) obtained using 1811 eigenfrequencies of the graph. The numerical NNSD is compared to the Poisson (green dashed line) and GOE (blue dash-dotted line) theoretical distributions. The fit of the formula (\ref{Eq.5}) (solid red line) to the numerical data yields the parameter $\rho_1=0.496 \pm 0.019$.  (d) The numerical spectral rigidity $\Delta_3(L)$ (grey open circles) is compared to the Berry-Robnik (solid red line), Poisson (green dashed line), and GOE (blue dash-dotted line) distributions. The error bars in panels (c) and (d) represent the standard deviation of uncertainty.
}\label{Fig5}	
\end{center}
\end{figure}

\begin{figure}[tb]
\begin{center}
\rotatebox{0}{\includegraphics[width=1.4\textwidth,
height=1.4\textheight, keepaspectratio]{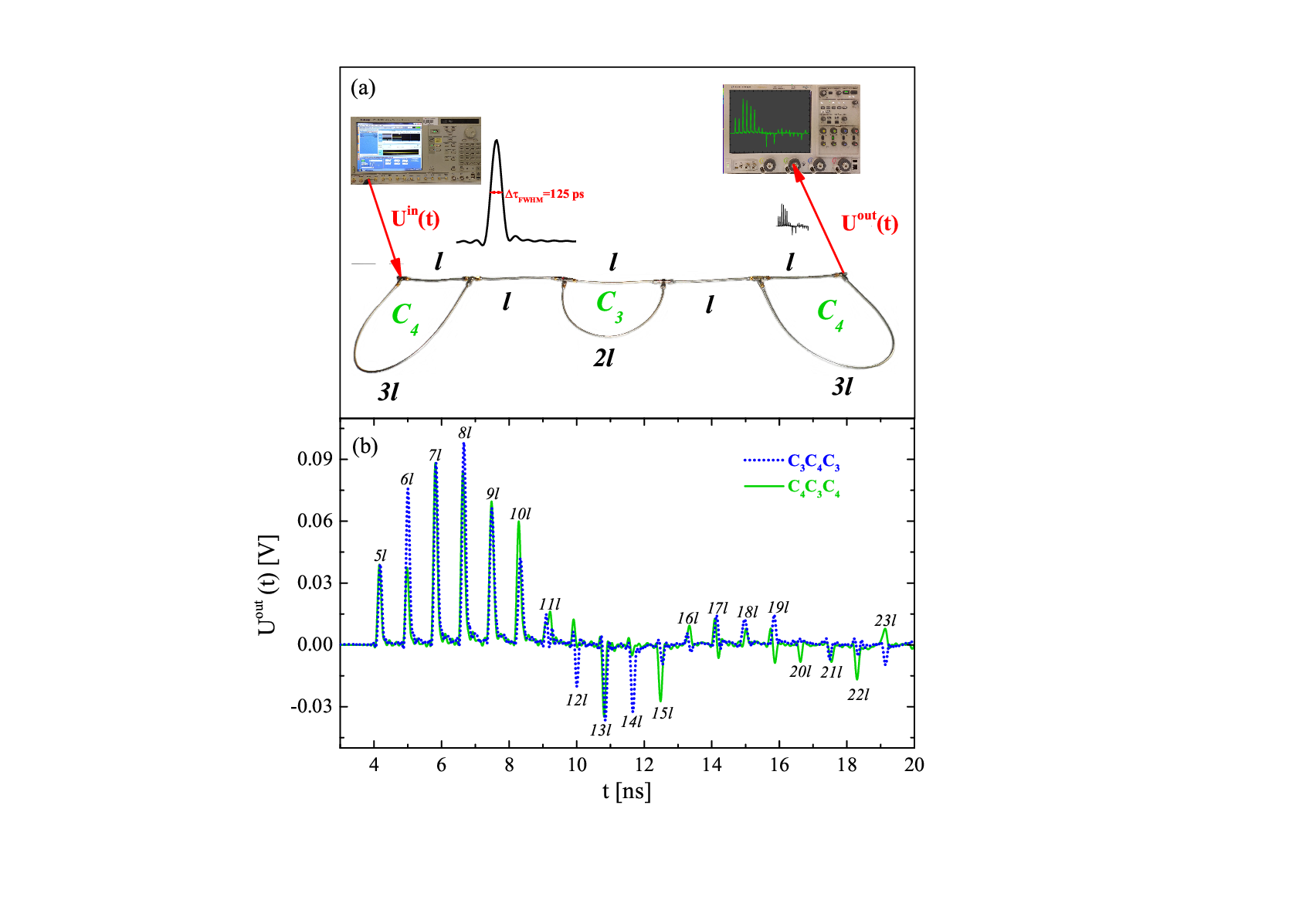}}
\caption{(a) The square-triangle-square $C_{4}C_{3}C_{4}$ microwave network simulating the quantum graph $C_{4}C_{3}C_{4}$. A Gaussian input signal
synthesized by the waveform generator AWG 7082C with the amplitude $A=0.41$ V and the FWHM of $125$ ps enters the network through the attached input lead and propagates inside it until it leaves through the output lead. The output signal is detected and stored by the digital oscilloscope MSO-X-91304A. The lengths of the edges: $l=0.250\pm 0.001$ m, $2l=0.500\pm 0.002$ m, and $3l=0.750\pm 0.002$ m. (b) The experimental delay-time distributions obtained for the microwave networks $C_{3}C_{4}C_{3}$ and $C_{4}C_{3}C_{4}$ are denoted by blue dotted and green solid lines, respectively.  The positive and negative peaks transmitted through the networks are formed due to the interference of signals traveling along many very complicated available paths. The peaks marked as $5l$, $6l$, $7l$ \ldots  are appropriately attributed to the paths of the length: $5l$, $6l$, $7l$ \ldots
}\label{Fig6}	
\end{center}
\end{figure}


\section{References}
\begin{thebibliography}{99}
\bibliographystyle{unsrt}
\bibitem{Pauling} L. Pauling, J. Chem. Phys. 4, 673 (1936).
\bibitem{Sanchez1998} J. A. Sánchez-Gil, V. Freilikher, I. Yurkevich, and A. A. Maradudin, Phys. Rev. Lett. 80, 948 (1998).
\bibitem{Imry2002} Y. Imry, Introduction to Mesoscopic Systems (Oxford University Press, New York, 2002).
\bibitem{Exner88} P. Exner, P. \v{S}eba, and P. \v{S}\v{t}ov\'i\v{c}ek, J. Phys. A {\bf 21}, 4009 (1988).
\bibitem{Kottos1999} T. Kottos and U. Smilansky, Ann. Phys. {\bf 274}, 76 (1999).
\bibitem{Blumel2002} R. Bl\"umel, Yu Dabaghian, and R. V. Jensen, Phys. Rev. Lett. {\bf 88}, 044101 (2002).
\bibitem{Hul2004} O.~Hul, S.~Bauch, P.~Pako\'nski, N.~Savytskyy, K.~{\.Z}yczkowski, and L.~Sirko, Phys. Rev. E {\bf 69}, 056205 (2004).
\bibitem{Berkolaiko2012} G. Berkolaiko, P. Kuchment, Introduction to Quantum Graphs (American Mathematical Society, Providence, (2012).
\bibitem{Kumar2013} S. Kumar, A. Nock, H.-J. Sommers, T. Guhr, B. Dietz, M. Miski-Oglu, A. Richter, and F. Sch\"afer,
Phys. Rev. Lett. {\bf 111}, 030403 (2013).
\bibitem{Lawniczak2020b} M. {\L}awniczak, B. van Tiggelen, and L. Sirko, Phys. Rev. E {\bf 102}, 052214 (2020).
\bibitem{Pluhar2014} Z. Pluha\v r and H. A. Weidenm\"uller, Phys. Rev. Lett. {\bf 112}, 144102 (2014).
\bibitem{Dietz2017} B. Dietz, V. Yunko, M. Bia{\l}ous, S. Bauch, M. {\L}awniczak, and L. Sirko,  Phys. Rev. E {\bf 95}, 052202 (2017).
\bibitem{Pinheiro2020} L. K. Pinheiro, B. S. Souza, V. Trevisian, Discussiones Mathematicae - Graph Theory {\bf 40}, 607-620, (2020).
\bibitem{Lawrie2022} T. Lawrie, G. Tanner, D. Chronopoulas, Scientific Reports 12, 18006 (2022).
\bibitem{Lawniczak2008} M.~{\L}awniczak, O.~Hul, S.~Bauch, P.~\v Seba, and L.~Sirko, Phys. Rev. E {\bf 77}, 056210 (2008).
\bibitem{Hul2012} O. Hul, M.~{\L}awniczak, S. Bauch, A. Sawicki, M. Ku\'s, and L. Sirko, Phys. Rev. Lett {\bf 109}, 040402 (2012).
\bibitem{Sirko2016} M.~{\L}awniczak, S. Bauch, and L. Sirko, in Handbook of Applications of Chaos Theory, eds. Christos Skiadas and Charilaos Skiadas (CRC Press, Boca Raton, USA, 2016), p. 559.
\bibitem{Lawniczak2019b} M.~{\L}awniczak and L. Sirko, Scientific Reports {\bf 9}, 5630 (2019).
\bibitem{Yunko2020} V. Yunko, M. Bia{\l}ous, and L. Sirko, Phys. Rev. E {\bf 102}, 012210 (2020).
\bibitem{Hul2005b} O.~Hul, O.~Tymoshchuk, S.~Bauch, P.M.~Koch, and L.~Sirko, J. Phys. A {\bf 38}, 10489 (2005).
\bibitem{LipovskyJPhysA2016} J. Lipovsk\'{y}, J. Phys. A {\bf 49}, 375202 (2016).
\bibitem{Lawniczak2019} M.~{\L}awniczak, J. Lipovsk\'{y}, and L. Sirko, Phys. Rev. Lett. {\bf 122}, 140503 (2019).
\bibitem{Lawniczak2020} M.~{\L}awniczak, P. Kurasov, Sz. Bauch, M. Bia{\l}ous, V. Yunko, and L. Sirko, Phys. Rev. E {\bf 101}, 052320 (2020).
\bibitem{Lawniczak2021c} M.~{\L}awniczak, P. Kurasov, S. Bauch, M. Bia{\l}ous, A. Akhshani, and  L. Sirko, Scientific Reports {\bf 11}, 15342 (2021).
\bibitem{Lawniczak2009} M.~{\L}awniczak, S. Bauch, O. Hul, and L. Sirko, Physica Scripta {\bf T135}, 014050 (2009).
\bibitem{Lawniczak2010} M.~{\L}awniczak, S.~Bauch, O.~Hul, and L.~Sirko, Phys. Rev. E {\bf 81}, 046204 (2010).
\bibitem{Lawniczak2012} M.~{\L}awniczak, S. Bauch, O. Hul, and L. Sirko, Physica Scripta {\bf T147}, 014018 (2012).
\bibitem{Allgaier2014} M. Allgaier, S. Gehler, S. Barkhofen, H.-J. St\"ockmann, and U. Kuhl, Phys. Rev. E {\bf 89}, 022925 (2014).
\bibitem{Bialous2016} M. Bia{\l}ous, V. Yunko, S. Bauch, M. {\L}awniczak, B. Dietz, and L. Sirko, Phys. Rev. Lett. {\bf 117}, 144101 (2016).
\bibitem{Stockmann2016} A. Rehemanjiang, M. Allgaier, C.H. Joyner, S. M\"uller, M. Sieber, U. Kuhl, and H.-J. St\"ockmann, Phys. Rev. Lett. {\bf 117}, 064101 (2016).
\bibitem{Dietz2021} J. Che, J. Lu, X. Zhang, B. Dietz, and G. Chai, Phys. Rev. E {\bf 103}, 042212 (2021).
\bibitem{Lawniczak2023} M. {\L}awniczak, A. Akhshani, O. Farooq, M. Bia{\l}ous, S. Bauch, B. Dietz, and L. Sirko, Phys. Rev. E {\bf 107}, 024203 (2023).
\bibitem{Dietz2010} B. Dietz, T. Friedrich, H. L. Harney, M. Miski-Oglu, A. Richter, F. Sch{\"a}fer, and H. A. Weidenm{\"u}ller, Phys. Rev. E {\bf 81}, 036205 (2010).
\bibitem{Yeh2013} J. -H. Yeh, Z. Drikas, J. Gil Gil, S. Hong, B. T. Taddese, E. Ott, T.M. Antonsen, T. Andreadis, and S.M. Anlage, Acta Phys. Pol. A {\bf 124}, 1045 (2013).
\bibitem{Zheng2006} X. Zheng, S. Hemmady, T. M. Antonsen, Jr. S. M. Anlage, and E. Ott, Phys. Rev. E {\bf 73} 046208 (2006).
\bibitem{Stockmann1990} H.-J. St\"ockmann and J. Stein, Phys. Rev. Lett. {\bf 64}, 2215 (1990).
\bibitem{Sridhar1994} S.~Sridhar and A.~Kudrolli, Phys. Rev. Lett. {\bf 72,} 2175 (1994).
\bibitem{Sirko1997} L. Sirko, P.M. Koch, and  R. Bl\"umel, Phys. Rev. Lett. {\bf 78}, 2940 (1997).
\bibitem{Hlushchuk2000} Y. Hlushchuk, A. Kohler, Sz. Bauch, L. Sirko, R. Bl\"umel, M. Barth, and H.-J. St\"ockmann, Phys. Rev. E {\bf 61}, 366-369 (2000).
\bibitem{Hlushchuk2001} Y. Hlushchuk, A. B{\l}\c{e}dowski, N. Savytskyy, L. Sirko, Physica Scripta {\bf 64}, 192 (2001).
\bibitem{Hlushchuk2001b} Y. Hlushchuk, L. Sirko, U. Kuhl, M. Barth, and H.-J. St\"ockmann Phys. Rev. E {\bf 63}, 046208 (2001).

\bibitem{Dhar2003} A. Dhar, D. M. Rao, U. Shankar, and S. Sridhar, Phys. Rev. E {\bf 68,} 026208 (2003).
\bibitem{Savytskyy2004} N. Savytskyy, O. Hul, and L. Sirko, Phys. Rev. E {\bf70}, 056209 (2004).
\bibitem{HemmadyPRL2005} S. Hemmady, X. Zheng, E. Ott, T.M. Antonsen, S.M. Anlage, Phys. Rev. Lett. {\bf 94}, 014102 (2005).
\bibitem{Hul2005} O. Hul, N. Savytskyy, O. Tymoshchuk, S. Bauch, and L. Sirko, Phys. Rev. E {\bf 72}, 066212 (2005).
\bibitem{Dietz2015} B. Dietz and A. Richter, CHAOS {\bf 25}, 097601 (2015).
\bibitem{Dietz2019} B. Dietz, T. Klaus, M. Miski-Oglu, A. Richter, and M. Wunderle, Phys. Rev. Lett {\bf123}, 174101 (2019).
\bibitem{Jensen1991}  R. V. Jensen, S. M. Susskind, and M. M. Sanders, Physics Reports {\bf 201}, 1 (1991).
\bibitem{Bellerman1992} M. Bellermann, T. Bergemann, A. Haffmann, P. M. Koch, and L. Sirko, Phys. Rev. A {\bf 46}, 5836 (1992).

\bibitem{Buchleitner1993} A. Buchleitner and D. Delande, Phys. Rev. Lett. {\bf 71}, 3633 (1993).
\bibitem{SirkoPRL1993} L. Sirko, M. R. W. Bellermann, A. Haffmans, P. M. Koch, and D. Richards, Phys. Rev. Lett. {\bf 71}, 2895-98 (1993).
\bibitem{Bayfield1995} J. E. Bayfield, S.-Y. Luie, L. C. Perotti, and M. P. Skrzypkowski, Physica D: Nonlinear Phenomena {\bf 83}, 46 (1995).
\bibitem{Sirko1995} L. Sirko and P. M. Koch, Appl. Phys. B {\bf 60}, S195 (1995).
\bibitem{Sirko1996} L. Sirko, A. Haffmans, M. R. W. Bellermann, and P. M. Koch,  Europhysics Letters {\bf 33}, 181 (1996).
\bibitem{Bayfield1999} J. Bayfield and Lal Pinnaduwage, J. Phys. B {\bf 18}, L49 (1999).
\bibitem{Sirko2001} L. Sirko, S. A. Zelazny, and P. M. Koch, Phys. Rev. Lett. {\bf 87}, 043002 (2001).

\bibitem{Galagher2016} A. Arakelyan, J.  Nunkaew, and T.F. Gallagher, Phys. Rev. A {\bf 94}, 053416 (2016).

\bibitem{Anlage2020}  L. Chen, T. Kottos, S.M. Anlage, Nature Communications {\bf 11},  5826 (2020).
\bibitem{Yusupov2019} J.R. Yusupov, K.K. Sabirov, M. Ehrhardt, D.U. Matrasulov, Physics Letters A {\bf 383}, 2382 (2019).

\bibitem{Schmidt2003} A. G. M. Schmidt, B. K. Cheng, and M. G. E. da Luz, J. Phys. A {\bf 36}, L545 (2003).
\bibitem{Andrade2016} F. M. Andrade, A. G. M. Schmidt, E. Vicentini, B. K. Cheng, and M. G. E. da Luz, Physics Reports {\bf 647}, 1 (2016).
\bibitem{Andrade2018} F. M. Andrade and S. Severini, Phys. Rev. A {\bf 98}, 062107 (2018).
\bibitem{Drinko2019} A. Drinko, F. M. Andrade and D. Bazeia, Phys. Rev. A {\bf 100}, 062117 (2019).
\bibitem{Drinko2020} A. Drinko, F. M. Andrade and D. Bazeia, Eur. Phys. J. Plus {\bf 135}, 451 (2020).
\bibitem{Estarellas2015} C. Estarellas and L. Serra, Superlattices and
Microstructures {\bf 83}, 184 (2015).

\bibitem{Farooq2022} O. Farooq,  M. {\L}awniczak, A. Akhshani, S. Bauch, and L. Sirko, Entropy {\bf 24}, 387 (2022).


\bibitem{BHJ} R. Band, J. M. Harrison, C. H. Joyner, J. Phys. A Math. Theor. {\bf 45}, 325204 (2012).
\bibitem{Li7} J. Lipovsk\'{y}, Acta Physica Polonica A, {\bf 128}, 968 -- 973 (2015).
\bibitem{Goubau1961} G. ~Goubau, Electromagnetic Waveguides and Cavities (Pergamon Press, Oxford, 1961).
\bibitem{Kottos2003} T. Kottos and U. Smilansky, J. Phys. A: Math. Gen. {\bf 36}, 350-1 (2003).
\bibitem{Li2021} J. Li, T. Prosen, and A. Chan, Phys. Rev. Lett. {\bf 127}, 170602 (2021).

\bibitem{Jones1964} D. S. Jones, {\it Theory of the Electromagnetism} (Pergamon Press, Oxford, 1964) p. 254
\bibitem{Savytskyy2001} N. Savytskyy, A. Kohler, Sz. Bauch, R. Bl\"{u}mel, L. Sirko, Phys. Rev. E {\bf 64}, 036211 (2001).
\bibitem{LawniczakAPPA2019} M. {\L}awniczak, S. Bauch, V. Yunko, M. Bia{\l}ous, J. Wrochna and L. Sirko, Acta Phys. Pol. A {\bf 136}, 811 (2019).

\bibitem{Mehta1990} M. L. Mehta, \textit{Random Matrices} (Academic, London, 1990).
\bibitem{Berry1984} M. V. Berry and M. Robnik, J. Phys. A: Math. Gen. {\bf 17}, 2413 (1984).
\bibitem{Prosen1994} T. Prosen and M. Robnik, J. Phys. A: Math. Gen. {\bf 27}, L459 (1994).
\end{thebibliography}
\end{document}